\documentclass[lettersize,journal]{IEEEtran}
\usepackage{amsmath,amsfonts}
\usepackage{algorithmic}
\usepackage{algorithm}
\usepackage{array}
\usepackage[caption=false,font=normalsize,labelfont=sf,textfont=sf]{subfig}
\usepackage{textcomp}
\usepackage{stfloats}
\usepackage{url}
\usepackage{verbatim}
\usepackage{graphicx}
\usepackage{cite}
\usepackage{CJKutf8}
\usepackage{booktabs}
\usepackage{graphicx}
\usepackage{multirow}
\usepackage{threeparttable}

\begin{document}
\begin{CJK}{UTF8}{gkai}

\title{Exploring the Role of Large Language Models in Cybersecurity: A Systematic Survey}

\author{Shuang Tian,
    Tao Zhang,~\IEEEmembership{Member,~IEEE,}
    Jiqiang Liu,~\IEEEmembership{Senior Member,~IEEE,}
    Jiacheng Wang, 
    Xuangou Wu,
    Xiaoqiang Zhu,~\IEEEmembership{Member,~IEEE,}
    Ruichen Zhang,~\IEEEmembership{Member,~IEEE,}
    Weiting Zhang,~\IEEEmembership{Member,~IEEE,}
    Zhenhui Yuan,~\IEEEmembership{Senior Member,~IEEE,}
    Shiwen Mao,~\IEEEmembership{Fellow,~IEEE,}
    Dong In Kim,~\IEEEmembership{Life Fellow,~IEEE}

\thanks{
This work was supported in part by the National Cryptologic Science Fund of China under Grant 2025NCSF02030; in part by the Talent Fund of Beijing Jiaotong University under Grant 2023XKRC050; in part by the National Natural Science Foundation of China (NSFC) under Grant 62402029; in part by the Open Fund of Anhui Provincial Key Laboratory of Digital Twin Technology in Metallurgical Industry under Grant ADW24-01; and in part by the China Postdoctoral Science Foundation under Grant 2024T170047, Grant GZC20230223, and Grant 2024M750165.
\textit{(Corresponding author: Tao Zhang and Weiting Zhang)}

S. Tian is with the School of Software Engineering, Beijing Jiaotong University, Beijing 100044, China. E-mail: tianshuang@bjtu.edu.cn.

T. Zhang, J. Liu and X. Zhu are with the School of Cyberspace Science
and Technology, Beijing Jiaotong University, Beijing 100044, China. E-mail: taozh@bjtu.edu.cn; jqliu@bjtu.edu.cn; xqzhu@bjtu.edu.cn.

J. Wang is with the School of Computer Science and Engineering, Nanyang Technological University, Singapore 639798. E-mail: jiacheng.wang@ntu.edu.sg.

X. Wu is with the School of Computer Science and Technology, An hui Province Key Laboratory of Digital Twin Technology in Metallurgical Industry, Anhui University of Technology, Anhui 243002, China. E-mail: wuxgou@ahut.edu.cn.

R. Zhang is with the College of Computing and Data Science, Nanyang Technological University, Singapore. E-mail: ruichen.zhang@ntu.edu.sg.

W. Zhang is with the School of Electronic and Information Engineering, Beijing Jiaotong University, Beijing 100044, China. E-mail: wtzhang@bjtu.edu.cn.

Z. Yuan is with the School of Engineering, University of Warwick, CV4 7AL Coventry, U.K. E-mail: zhenhui.yuan@warwick.ac.uk.

S. Mao is Department of Electrical and Computer Engineering, Auburn University, Auburn, AL 36849-5201, USA. E-mail: smao@auburn.edu.

D. I. Kim is with the Department of Electrical and Computer Engineering, College of Information and Communication Engineering, Sungkyunkwan University, Suwon 16419, South Korea. E-mail: dongin@skku.edu.
}}

\markboth{ IEEE TRANSACTIONS ON NETWORK SCIENCE AND ENGINEERING,~Vol.~XX, No.~XX, XXXX~XXXX}%
{Shell \MakeLowercase{\textit{et al.}}: A Sample Article Using IEEEtran.cls for IEEE Journals}

\IEEEpubid{0000--0000/00\$00.00~\copyright~2021 IEEE}

\maketitle

\begin{abstract}

With the rapid development of technology and the acceleration of digitalisation, the frequency and complexity of cyber security threats are increasing. Traditional cybersecurity approaches, often based on static rules and predefined scenarios, are struggling to adapt to the rapidly evolving nature of modern cyberattacks. There is an urgent need for more adaptive and intelligent defence strategies. The emergence of Large Language Model (LLM) provides an innovative solution to cope with the increasingly severe cyber threats, and its potential in analysing complex attack patterns, predicting threats and assisting real-time response has attracted a lot of attention in the field of cybersecurity, and exploring how to effectively use LLM to defend against cyberattacks has become a hot topic in the current research field. This survey examines the applications of LLM from the perspective of the cyber attack lifecycle, focusing on the three phases of defense reconnaissance, foothold establishment, and lateral movement, and it analyzes the potential of LLMs in Cyber Threat Intelligence (CTI) tasks. Meanwhile, we investigate how LLM-based security solutions are deployed and applied in different network scenarios. It also summarizes the internal and external risk issues faced by LLM during its application. Finally, this survey also points out the facing risk issues and possible future research directions in this domain.
\end{abstract}

\begin{IEEEkeywords}
Large Language Model, Cybersecurity, Cyber attacks, Cyber defense, Intrusion detection, Anomaly detection， Phishing attack detection, Malware detection, Vulnerability detection, Vulnerability patch, Cyber Threat Intelligence, next-generation networks
\end{IEEEkeywords}

\section{Introduction}
\label{section_1}



\IEEEPARstart{A}{s} the information age develops rapidly, cyberattacks are taking on the characters of high frequency, diversity and complexity \cite{bout2021machine, liu2021robust, chen2024survey, wang2024generative}. Critical infrastructure and personal sensitive data are facing a broad range of novel threats, including malware, ransomware, and DDoS attacks \cite{syrmakesis2023novel, kan2021novel, yazdinejad2023secure}. The evolution of threat methods and the growing intensity of these threats is resulting in severe economic and property damage \cite{soikkeli2022redundancy}. In the United States, a ransomware attack on Colonial Pipeline system completely shuttered their operations, leading to a gasoline shortage across the country's East Coast for an entire week. In 2022, Sunwing Airlines were forced to cancel 188 flights and leave passengers stranded at airports for over three days, all due to a cyber attack on their supplier’s systems. These are just a few of an ever-growing list of cyberattacks that are altering different spheres of daily life, making cybersecurity one of the core foundational issues of modern global security concern \cite{aldaajeh2022role}.


As cyberattacks continue to evolve in their persistence, stealth, and unpredictability, existing cybersecurity measures struggle to keep pace in detecting, preventing and mitigating threats to networks. Traditional network defence methods based on fixed rules and scenarios have been exhausted in the complex network environment \cite{ding2024large}. Although advanced AI-based cyber defence methods have developed rapidly in recent years with the rise of neural networks and deep learning technologies. However, these methods suffer from high false positives and lack of interpretability when put into commercial use \cite{taddeo2019trusting}, making it still a huge challenge to deal with rapidly evolving forms of cyber-attacks. Additionally, cybersecurity researchers are confronted with numerous challenges due to the performance limitations of these systems, including managing large volumes of sensitive data and dealing with the complexities of cybersecurity tasks\cite{chen2024survey}. In response to this situation, cybersecurity researchers recognise the need for stronger, more adaptable and smarter solutions to deal with the ever-increasing threat of attacks.


\IEEEpubidadjcol

The emergence of LLM has provided a new way of thinking about network defence. In recent years, LLMs have achieved significant breakthroughs in the field of natural language processing and have also shown great potential in cybersecurity defense. By leveraging extensive training datasets, LLMs can identify latent attack patterns and vulnerabilities, assist in analyzing attack behaviors, predict threats, and even provide real-time defensive support. They are capable of identifying cybersecurity risks based on historical attack data or contextual attack information and can proactively generate response strategies. 
LLM is used for a variety of cybersecurity tasks, such as threat detection, analysis of cybersecurity reports and the provision of defence recommendations.


In view of the substantial value of LLMs currently show in cybersecurity, we would like to provide an overview of present-day applications of LLMs in this domain, in order to provide future researchers with an outlook and ideas. In summary, the contributions made in this article are as follows:
\begin{itemize}
\item{The feasibility and application prospects of LLMs in cybersecurity are discussed in depth through a systematic investigation of current benchmarking studies that assess the performance of LLMs and specific technical means to optimise the behaviour of LLMs in cybersecurity tasks.}
\item{An innovative and comprehensive analysis of the defensive role played by LLM from the attacker's point of view across the various lifecycles of a cyberattack. Additionally, due to the important intelligence base role played by CTI in defence operations, and as a complement, we have also explored the defence role played by LLMs in CTI work.}
\item{We also analyze the deployment and application approaches of LLM-based security applications in different network scenarios and the challenges they face}
\item{This paper analyses the external and internal risks that LLMs may face in the process of executing cybersecurity tasks and provides risk warning and coping ideas for related research and applications.}
\end{itemize}


\begin{figure*}[!t]
\centering
\includegraphics[width=1.0\textwidth]{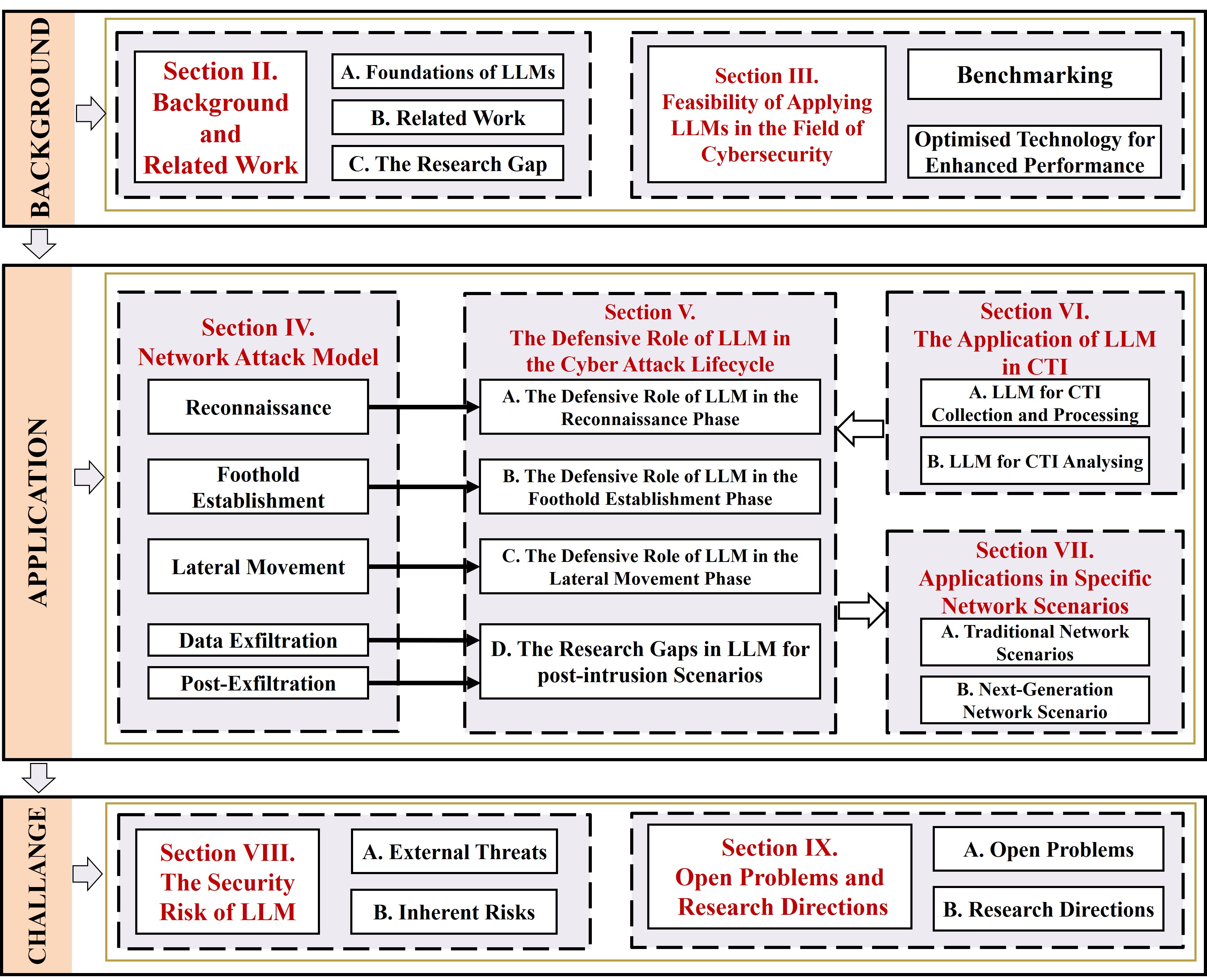}
\caption{The overall organizational structure of this survey.}
\label{structure}
\end{figure*}

The organizational structure of this article is illustrated in Fig. \ref{structure}. Section \ref{section_2} introduces the foundations of the LLM and provides a concise survey of existing investigations into the application of LLMs for cybersecurity, while also identifying and analyzing current research gaps. Section \ref{section_3} briefly reviews research evaluating the performance of LLM in cybersecurity and optimising LLM technology in this field. Section \ref{section_4} introduces the network attack model employed. 
Sections \ref{section_5} explores the applications and limitations of LLMs in different phases of a network attack lifecycle. Section \ref{section_6} explores the applications of LLMs in CTI. Section \ref{section_7} explores the deployment and application of LLM-based network security solutions in different network scenarios. Section \ref{section_8} summarizes the internal and external risks associated with applying LLMs in the network security domain. Section \ref{section_9} summarises the current challenges encountered by LLMs in cybersecurity tasks and future research directions in this domain. Section \ref{section_10} summarises our research. Table \ref{acronyms table} lists and describes the acronyms used throughout this paper.

\begin{table}
\begin{center}
\renewcommand{\arraystretch}{1.1}
\caption{List of Acronyms Used Throughout This Paper.}
\label{acronyms table}
\begin{tabular}{m{1.5cm} m{6cm}}
\toprule
\textbf{Acronym} & \textbf{Definition} \\
\midrule
LLM & Large Language Model \\
CTI & Cyber Threat Intelligence \\
IDS & Intrusion Detection Systems \\
NER & Named Entity Recognition \\
ICS & Industrial Control Systems \\
PEFT & Parameter-Efficient Fine-Tuning \\
EDR & Endpoint Detection and Response \\
MHA & Multi-Head Attention \\
CoT & Chain-of-Thought \\
PbE & Programming by Example \\
EMAD & Evidence-Based Multi-Agent Debate \\
GNN & Graph Neural Network \\
RAG & Retrieval-Augmented Generation \\
CEC & Contract-External Function-Call \\
IoT & Internet of Things \\
USE & Unidirectional Semantic Extractor \\
BSE & Bidirectional Semantic Extractor \\
ML & machine learning \\
BS & base station \\
NF & network function \\
LEGD & Large Language Model-Enhanced Graph Diffusion \\
DTN & digital twins network \\
SAGIN & satellite-aerial-ground integrated network \\
SLM & Small Language Model \\
\bottomrule
\end{tabular}
\end{center}
\end{table}

\section{Background and related work}
\label{section_2}


In this section, we will briefly introduce the knowledge about LLM that is required to read this paper, and then review the relevant review literature on the current research on the application of LLM in cybersecurity in order to present the Research gap that currently exists.

\subsection{Foundations of LLMs}


As a cutting-edge technology in the field of artificial intelligence, LLM has been widely used in many fields and has become one of the current research hotspots. Its core technology is built on the ransformer architecture \cite{chen2024survey}, which can be divided into three typical structures based on decoding strategies \cite{kheddar2024transformers}:
\begin{itemize}
    \item The encoder-only architecture is excellent at language comprehension tasks, and is usually used for linguistic feature extraction, and the BERT model is a representative model of this architecture.
    \item The encoder-Decoder architecture is widely used in sequence-to-sequence tasks, and is widely used in text translation and speech recognition.
    \item The decoder-only architecture is the current research hot architecture, and the popular GPT series is based on this architecture.
\end{itemize}
After pre-training with large-scale datasets, LLMs can acquire language comprehension and logical reasoning abilities, and if trained with domain-specific datasets during the pre-training process, LLMs can even perform comparably to humans in some domains. In addition, fine-tuning and prompt engineering techniques are important complementary means to optimise the model's ability to perform domain-specific tasks in the post-pretraining phase, and the use of these two techniques can effectively expand the scope of application of pre-trained models. However, it should be noted that there are still some challenges that hinder the application scope and performance of LLM, such as high quality domain training sets, high training costs and inference delay issues.

\subsection{Related Work}

\begin{table*}
\begin{center}
\renewcommand{\arraystretch}{1.3}
\caption{Comparison of LLM Application Surveys in Cybersecurity. ``●” and ``○” Represent Rxplored and Not Rxplored, Respectively.}
\label{comparison table}
\begin{tabular}{p{2.2cm} >{\raggedright\arraybackslash}p{2cm} >{\raggedright\arraybackslash}p{2.7cm} >{\raggedright\arraybackslash}p{2.2cm} >{\raggedright\arraybackslash} >{\raggedright\arraybackslash}p{1.9cm} >{\raggedright\arraybackslash}p{2.3cm} >{\raggedright\arraybackslash}p{1.7cm}}
\toprule
\textbf{Surveys} & \textbf{Defense against Reconnaissance Attack} & \textbf{Defence against Foothold Establishment Attack} & \textbf{Defence against Lateral Movement Attack} & \textbf{Cyber Threat Intelligence Work} & \textbf{Application in Different Network Scenarios} & \textbf{LLM's Own Security Risks} \\
\midrule
Kheddar \cite{kheddar2024transformers} & ○ & ○ & ● & ○ & ● & ●
\\
Sheng \textit{et al.} \cite{sheng2025large} & ○ & ● & ○ & ○ & ○ & ○
\\
Zhou \textit{et al.} \cite{zhou2024largereview} & ○ & ● & ○ & ○ & ● & ○
\\
Zhang \textit{et al.} \cite{zhang2024llms} & ○ & ● & ● & ● & ○ & ●
\\
Hang \textit{et al.} \cite{hang2024large} & ○ & ● & ● & ● & ○ & ○
\\
Motlagh \textit{et al.} \cite{motlagh2024large} & ● & ● & ● & ○ & ○ & ○
\\
Chen \textit{et al.} \cite{chen2024survey} & ● & ● & ● & ● & ○ & ○
\\
Yao \textit{et al.} \cite{yao2024survey} & ○ & ● & ○ & ○ & ○ & ●
\\
\textbf{Our survey} & ● & ● & ● & ● & ● & ●
\\
\bottomrule
\end{tabular}
\end{center}
\end{table*}


Current review studies on LLM in network security usually focus on the application methods, usage scenarios, and performance evaluation of LLM. In this subsection, we review these studies and describe the contributions and insights they provide.

\subsubsection{Focused Analysis of LLMs in Specific Cybersecurity Tasks}


Some surveys have focused its research on the performance of LLMs in specific cybersecurity tasks in order to deeply analyse their capabilities and limitations in practical applications.
For instance, Ref. \cite{kheddar2024transformers} systematically combs through the research on LLM-based intrusion detection systems (IDSs) in different architectures and deployment environments while demonstrating their practical utility through real-world use cases. However, the article also points out that LLM-based IDS systems still face many challenges, such as training data privacy issues, network data heterogeneity, and inherent security vulnerabilities in LLM architectures. 
Ref. \cite{sheng2025large} investigates the current state of the application of LLMs in vulnerability detection. The authors point out that the dominant model architecture in the field is shifting from encoder-only to decoder-only. In addition, existing research is overly reliant on C/C++ language vulnerability datasets and lacks repository-level data, which limits LLM's ability to generalise across languages and detect complex multi-file vulnerabilities.
Ref. \cite{zhou2024largereview} summarises the main types of LLMs and the main LLM performance optimisation techniques in vulnerability detection and repair. The article also points out that there is a lack of class-level or repository-level training datasets in this area, the reliability of existing datasets is poor, and the datasets often lack test samples. Finally, the authors argue that much of the current research does not emphasise integration with developer workflows, and that there is a lack of mechanisms for interaction between users and LLMs.

\subsubsection{Broad Exploration of LLMs in Multi-Domain Cybersecurity Applications}


Some surveys have taken a broader perspective to deeply analyze the performance of LLMs across multiple key tasks for cybersecurity.
Both Ref. \cite{zhang2024llms} and Ref. \cite{hang2024large} provide systematic summaries and organization of current research on the application of LLMs in cybersecurity. While the two surveys unanimously acknowledge that LLMs can significantly improve the efficiency of cybersecurity tasks, they also highlight persistent challenges, including external attack threats and inherent limitations in model performance. 
Within the the National Institute of Standards and Technology Cybersecurity Framework, Ref. \cite{motlagh2024large} studies LLM applications in the identify, protect, detect, respond, and recover phases. It notes that current research focuses on protect and detect scenarios, but post-attack scenarios, including response and recovery phases, remain understudied. Given their critical roles, expanding LLM research in these areas is essential for comprehensive cybersecurity.
Ref. \cite{chen2024survey} examines LLM applications in four key threat detection areas: CTI, textual threat detection, malware detection, and intrusion discovery. It reveals that LLMs surpass traditional methods primarily in specific tasks like NER, Relation Extraction, and structured information processing. However, for significantly more sophisticated threat detection scenarios, LLMs typically necessitate integration with complementary technologies for optimal performance.
Ref. \cite{yao2024survey} examines the application of LLMs in code security tasks, such as secure coding and vulnerability detection. The survey reveals that LLM-based approaches generally surpass traditional methods in this domain, although they exhibit higher rates of both false negatives and false positives. Furthermore, through an investigation of LLMs' application in data security tasks—encompassing data integrity, confidentiality, and reliability—the research demonstrates that LLMs not only minimize manual intervention but also achieve superior performance in these areas. Both Ref. \cite{zhang2024llms} and Ref. \cite{hang2024large} provide systematic summaries and organization of current research on the application of LLMs in cybersecurity. While the two surveys unanimously acknowledge that LLMs can significantly improve the efficiency of cybersecurity tasks, they also highlight persistent challenges, including external attack threats and inherent limitations in model performance. 
Within the the National Institute of Standards and Technology Cybersecurity Framework, Ref. \cite{motlagh2024large} studies LLM applications across identify, protect, detect, and respond stages. While current research predominantly focuses on Protect and Detect, post-attack scenarios, encompassing the Respond and Recovery phases, remain significantly understudied. Given their important role, expanding LLM research in these areas is essential for comprehensive cybersecurity.
Ref. \cite{chen2024survey} examines LLM applications in four key threat detection areas: Cyber Threat Intelligence (CTI), textual threat detection, malware detection, and intrusion discovery. It reveals that LLMs surpass traditional methods primarily in specific tasks like NER, Relation Extraction, and structured information processing. However, for significantly more sophisticated threat detection scenarios, LLMs typically necessitate integration with complementary technologies for optimal performance.
Ref. \cite{yao2024survey} examines the application of LLMs in code security tasks, such as secure coding and vulnerability detection. The survey reveals that LLM-based approaches generally surpass traditional methods in this domain, although they exhibit higher rates of both false negatives and false positives. Furthermore, through an investigation of LLMs' application in data security tasks—encompassing data integrity, confidentiality, and reliability—the research demonstrates that LLMs not only minimize manual intervention but also achieve superior performance in these areas.

\subsection{The Research Gap}

The research of the existing review papers in this section focuses on evaluating the performance of LLMs in specific tasks or application scenarios. However, for the systematic defence role of LLM in the whole network attack and defence process, there is still a certain research gap in current research. This gap has resulted in an incomplete understanding of the overall efficacy of LLMs in more holistic and dynamic cyber defense contexts. In addition, as a key intelligence component in cybersecurity defence, current research on CTI usually evaluates the performance of LLMs in isolated CTI tasks and lacks the defence role that LLMs can play from the whole CTI lifecycle. On the other hand, most studies nowadays have also overlooked the deployment of LLM in applications, and the lack of research here may create obstacles for future real-world applications. The surveys covered in this section are shown in Table \ref{comparison table}.

Consequently, this survey seeks to bridge this gap by adopting an innovative perspective rooted in the attacker lifecycle. It systematically and comprehensively examines the defensive role of LLMs at various stages of the cyber attack lifecycle. Additionally, it evaluates their performance in different stages of CTI tasks for effective integration with real-world defense actions.  Also investigating the way in which LLM-based security solutions are deployed at the time of application. This comprehensive analysis not only enhances the understanding of the practical value of LLMs in real-world cyber defense environments but also offers solid theoretical support for further research and the execution of defense measures.

\section{Feasibility of Applying LLMs in the Field of Cybersecurity}
\label{section_3}

Although LLMs are widely used in many fields \cite{shu2024LawLLM, Qiu2024LLMbase, wen2024AI}, due to the highly professional and complex of cybersecurity tasks, it is still questionable whether LLMs can perform these tasks efficiently. To explore this issue, researchers evaluated the performance of LLMs in cybersecurity tasks through benchmarking and further explored techniques to optimise the performance of LLMs in cybersecurity tasks.


Liu \cite{liu2023secqa} introduced SecQA, a benchmarking tool designed to evaluate LLM performance in computer security. Liu used GPT-4 to generate two multiple-choice question sets, v1 and v2, based on the content of a computer security book. v1 focused on LLM's basic understanding and application of cybersecurity knowledge, while v2 examined LLM's more in-depth and comprehensive understanding of advanced security topics through the use of more complex and detailed questions. Through experimental evaluation, the results show that GPT-3.5-Turbo and GPT-4 maintain high accuracy rates on the v2 set.
Tihanyi \textit{et al.} \cite{tihanyi2024cybermetric} constructed the CyberMetric-80 dataset to evaluate LLMs' cybersecurity knowledge coverage, which underwent rigorous expert validation to ensure answer accuracy. In a controlled evaluation, multiple LLMs and human participants completed the CyberMetric-80 assessment. The findings indicated that LLMs, especially GPT-4o and GPT-4-turbo, exhibited expertise comparable to seasoned cybersecurity professionals.
The results of these two benchmark tests suggest that LLMs have a strong foundation in cybersecurity knowledge and can effectively comprehend and apply it.

Liu \textit{et al.} \cite{liu2024cyberbench} evaluated generative LLMs in cybersecurity using a multi-task framework benchmarking with 10 datasets corresponding to four representative security tasks: NER, summarization, multiple choice, and text classification.
Bhusal \textit{et al.} \cite{bhusal2024secure} proposed the SECURE benchmarking framework that can be used to assess the capabilities of LLMs in industrial control systems (ICS) security consulting within three key abilities:
\begin{itemize}
\item{\textbf{Extraction}: Evaluates the efficiency of information retrieval using datasets from MITRE ATT\&CK and CWE.}
\item{\textbf{Understanding}: Employs the Vulnerability Out-Of-Distribution test set to determine whether models can identify unanswerable questions in the absence of contextual information.}
\item{\textbf{Reasoning}: Uses the Risk Evaluation Reasoning Task, constructed from CISA ICS security reports, to assess models' reasoning abilities in risk evaluation.}
\end{itemize}
Using the two previously mentioned benchmark frameworks, multiple LLMs were assessed, with models like GPT-4 scoring highly, indicating that LLMs still perform well on specific security tasks.

Although several studies have demonstrated the potential of LLMs for cybersecurity applications through benchmarking, most LLMs are not specifically designed for this domain task and may suffer from performance degradation due to lack of domain knowledge.This issue can be addressed with specialized techniques that can significantly improve their performance. Fine-tuning, prompt engineering, and domain-specific pre-training have been demonstrated to improve the performance of LLMs in this domain \cite{zhang2024llms}. For example, Zhang \textit{et al.} \cite{zhang2023hackmentor} developed tailored instructions and conversations for cybersecurity fine-tuning and applied LoRA fine-tuning to baseline LLMs, achieving a 10\%–25\% performance improvement. Similarly, Siracusano \textit{et al.} \cite{siracusano2023time} utilised specially designed prompts to guide their structured CTI extraction framework, aCTIon, thereby reducing the hallucinations when dealing with complex CTI data. In another study, Liu \textit{ et al.} \cite{liu2024cyberbench} introduced CyberDirective, a generative LLM that is fine-tuned on the CyberBench dataset using instruction tuning and parameter-efficient fine-tuning (PEFT), and demonstrated excellent in multiple cybersecurity tasks with excellent performance. The studies mentioned above highlight the efficacy of the  specialized techniques in enhancing LLM's domain-specific capabilities, particularly in cybersecurity applications.


Although the specialized and complex character of cybersecurity tasks challenges the applicability of LLM, many studies have shown that LLM already exists in a wide range of applications in several cybersecurity sub-domains and exhibits great potential for application. Meanwhile, the development of techniques such as fine-tuning, domain-specific pre-training, and prompt ngineering provides strong support for improving the performance of LLM in cybersecurity tasks. Results from various benchmark and performance tests also indicate that LLMs are increasingly capable in terms of understanding, reasoning, and knowledge coverage, gradually meeting the practical demands of cybersecurity work. Thus, it can be stated that the implementation of LLMs in  cybersecurity is progressing at a high pace and will be key role in improving productivity, automating analysis, and facilitating decision making in the future.

\section{Network Attack Model}
\label{section_4}

\begin{figure*}[!t]
\centering
\includegraphics[width=1\textwidth]{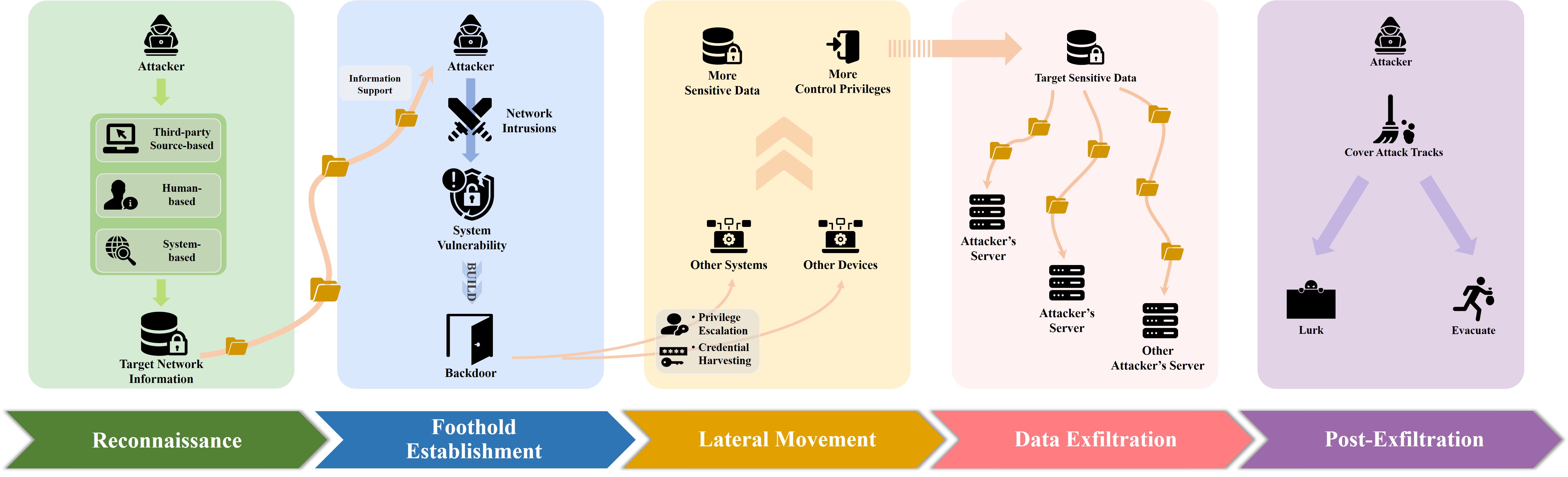}
\caption{Network attack model.}
\label{network attack model}
\end{figure*}


In order to easily describe the role played by LLM in each phase of cyber defence, it is necessary first to build a cyber attack model to describe the cyber attack process. In this paper, we divide the life cycle of an external cyber attack into five phases: reconnaissance, foothold establishment, lateral movement, data exfiltration, and post-exfiltration\cite{alshamrani2019survey}, as shown in Fig. \ref{network attack model}.

Reconnaissance refers to the process by which an attacker gathers information about the target network at the beginning of an attack. During this stage, attackers usually use covert passive reconnaissance methods \cite{alshamrani2019survey}. At this stage, attackers may collect information from public resources on the Internet, such as WHOIS websites \cite{sebastian2023domain} and the Google Hacking Database \cite{Saraswathi2022Automation}. Attackers may also use social engineering techniques to obtain information from users of the target network \cite{tian2022enhanced}, such as phishing attacks, enticing users of the target network to insert physical media with viruses into their computers, causing their computers to be infected with viruses \cite{tian2022enhanced}, and directly physically intruding such as tailgating attacks. In addition, attackers may also use technical methods to interact directly or indirectly with the target computer system to collect information, such as TCP scanning, ARP scanning, UDP scanning\cite{roy2022survey} to obtain information about the target system.


The foothold establishment phase signifies that the attacker has successfully infiltrated the target network. Attackers often create a ``foothold" within the target network to maintain long-term access to the target network system. Standard techniques include exploiting known vulnerabilities or zero-day vulnerabilities in web applications\cite{wang2020hybrid}, using spear-phishing attacks to implant a backdoor on the target's endpoint device\cite{bijmans2021catching}, or carrying out watering-hole attacks\cite{gan2024attack}, where attackers infect websites frequently visited by the target network's users, injecting malicious code into the victim's devices.


The lateral movement phase aims to expand the attacker's access to the target network to support the attacker in obtaining more sensitive data and control privileges of the target system. In this phase, the attacker usually uses Privilege Escalation, Credential Harvesting, and other means to expand the scope of his activities in the target network. In this phase, the techniques usually used are using the vulnerability of the target network system to enhance the attacker's privileges in the target system, using credentials dumping\cite{mohamed2021sbi}, hash transfer attacks\cite{al2021hidden} and session stealing techniques\cite{sharma2023advanced} to illegally obtain the user credentials of the target network to obtain a broader range of access privileges. Moreover, it has become challenging to completely eliminate the attacker from the system at this phase because the attacker has deep roots in the network and gained persistent control.


In the data exfiltration phase, the attacker transmits the stolen data to an external server under his control. In many network systems, firewalls and other security measures often focus on filtering incoming traffic, with little or no outbound traffic monitoring. This design facilitates data transfer, allowing attackers to easily pass sensitive information back. Typically, attackers will divide the collected data into multiple batches and send them to different servers in their organisation, which greatly reduces the possibility of the transmission being discovered by the defender.


The post-exfiltration phase is the action taken by the attacker after completing the attack's intended target. Typically, the attacker has two choices: the first is to evacuate the target network system and cover up the traces of the attack as much as possible, deleting any evidence left in the system; the second is to choose to continue to lurk in the target system in order to carry out new attacks in the future. This type of attacker usually hides their control privileges to maintain access to the target network for a long time without being detected.


There is a near linear flow relationship between the above five phases, but sometimes the attacker may also continue the previous phase of the task in parallel when proceeding to the next phase for reasons such as lack of preparation for the attack. For instance, the attacker may continue to carry out reconnaissance work while establishing a foothold.



\section{The Defensive Role of LLM in the Cyber Attack Lifecycle}
\label{section_5}


In this section, we will talk in detail about how LLM plays a defensive role in the various lifecycles of a network attack.

\subsection{The Defensive Role of LLM in the Reconnaissance Phase}


Reconnaissance attacks are the beginning stage of cyberattacks \cite{zhang2023disturb}, in which the attacker's reconnaissance behaviour is usually highly dispersed and covert, which poses a challenge to the defender's detection and blocking efforts. The reconnaissance attacks in this phase can be classified into three types: third-party source-based reconnaissance, which obtains attack information from third parties (e.g. third-party websites and dark web), human-based reconnaissance, which obtains attack information from the target network users, and system-based reconnaissance, which obtains attack information from the target computer system (hardware or software)\cite{roy2022survey}. LLM can effectively detect human-based and system-based reconnaissance attacks, helping network defenders better prevent the leakage of their sensitive information. All related works in this subsection are summarized in Table \ref{reconnaissance table}.

\begin{table*}
\begin{center}
\renewcommand{\arraystretch}{1.0}
\caption{Thematic Works about LLMs on Defense Against Reconnaissance Attacks.}
\label{reconnaissance table}
\begin{tabular}{m{2cm} m{3.5cm} m{5.5cm} m{2.5cm} m{1.2cm}}
\toprule
\textbf{Task} & \textbf{Detection Target} & \textbf{Detection Approach} & \textbf{Techniques} & \textbf{Literature} \\
\midrule
\multirow{3}{2 cm}{\parbox[c][2.5cm][c]{2cm}{System-based Reconnaissance Detection}} & Detect the advanced adversaries in \textbf{Smart Satellite Networks}. & \textbf{Transform network data} into contextually suitable inputs to \textbf{capture contexts and long-range relationships}. & Transformer-based MHA & \cite{hassanin2025pllm} \\ \cmidrule(l){2-5}
~ & Detect the \textbf{malicious user activities}. & Use \textbf{three self-designed agents} to detect attack behaviors by \textbf{detecting log files}. & CoT reasoning,\newline PbE paradigm,\newline EMAD mechanism & \cite{song2024audit} \\ \cmidrule(l){2-5}
~ & Attack \textbf{early detection}, threat intelligence gatherin, and analysis of attacker’s behavior. & Use \textbf{LLM-based honeypot} to simulate realistic shell responses and manages sessions for every attacker. & CoT prompting, Few-shot learning,\newline Session management. & \cite{sladic2024llm} \\ \midrule
\multirow{4}{2 cm}{\parbox[c][2.5cm][c]{2cm}{Human-based Reconnaissance Detection}} & Detect malicious web pages. & Use the question-and-answer \textbf{detection example} to guide the LLM. & K-means clustering,\newline Few-shot Prompting & \cite{li2023prompting} \\ \cmidrule(l){2-5}
~ & Detect \textbf{phishing email}. & Translate email into \textbf{LLM-readable} format and use \textbf{CoT} to guide the LLM. & Prompt Engineering,\newline CoT prompting & \cite{koide2024chatspamdetector} \\ \cmidrule(l){2-5}
~ & Detect \textbf{malicious behaviours} in the code. & Use prompt and Provide \textbf{complete contextual information}.  & Prompt Engineering & \cite{fang2024large} \\ \cmidrule(l){2-5}
~ & Detect and classify \textbf{malware}. & Use \textbf{few-shot} and \textbf{episodic training} to enhance LLM malware detection and classification. & Few-shot learning,\newline Episodic training & \cite{stein2024towards} \\
\bottomrule
\end{tabular}
\end{center}
\end{table*}

\subsubsection{LLM for System-based Reconnaissance Attack Detection}


In reconnaissance attacks targeting information systems, attackers frequently utilize remote scanning or sniffing to extract sensitive data from targeted systems. While these reconnaissance activities inevitably generate detectable traces within the target system, conventional detection approaches relying on rule-based or signature-based methodologies demonstrate limited effectiveness against sophisticated attack patterns. Conversely, LLMs, leveraging their advanced pattern recognition capabilities, demonstrate substantial advantages in both the precise identification and predictive analysis of reconnaissance attack behaviors.


Hassanin \textit{et al.} \cite{hassanin2025pllm} proposed a pre-trained LLM, PLLM-CS. The model initially generated sentences from the multivariate token series in the network traffic data. It then divided these sentences into tokens to capture contexts and long-range relationships within the traffic through the Transformer-based Multi-Head Attention (MHA) mechanism, which helped identify fragmented and frequent probing behaviors during attack detection. 
Song \textit{et al.} \cite{song2024audit} proposed Audit-LLM, a framework for detecting external attacks through log analysis, which consists of three agents: a Decomposer, a Tool Builder, and an Executor. The Decomposer uses Chain-of-Thought (CoT) reasoning to decompose complex tasks into subtasks. The Tool Generator generates Python tools using programming by example (PbE) paradigms, ensuring reliability through testing and refinement. The executor completes subtasks using CoT reasoning and employs paired evidence-based multi-agent debate (EMAD) mechanisms to reduce LLM illusions, iteratively optimizing results until consensus is reached.


In addition, there is now some research into using LLMs to create new types of honeypot systems. LLMs can use their knowledge base and memory capabilities to create deceptive system environments for different attackers to slow down the detection of the attackers.
Sladić \textit{et al.} \cite{sladic2024llm} proposed shelLM, a shell-based honeypot software using LLM. This honeypot uses the CoT prompting and few-shot learning and it can generates responses that are consistent with a real Linux shell based on the interaction history and the attacker's commands. Meanwhile, due to the non-deterministic nature of LLM, shelLM can simulate multi-user environments and enhance the realism of the honeypot system. Evaluations conducted by volunteers show a true negativity rate of 0.9, indicating that it can effectively mimic the responses of a real system and deceive users in 90\% of the cases.

\subsubsection{LLM for Human-based Reconnaissance Attack Detection}


Social engineering is a widely used technique in cyberattacks, often resulting in significant data breaches\cite{tian2022enhanced}. Attackers typically employ methods such as phishing and watering-hole attacks to deceive users and extract sensitive system information. Recent advancements in LLM technology have shown promising results in detecting phishing attacks and identifying malware, offering innovative technical solutions to combat these social engineering threats.


Malicious web pages and phishing are two common means of detecting attacks. 
Li \textit{et al.} \cite{li2023prompting} introduced Prompt-URL, a few-shot prompting approach for LLM-based malicious webpage detection. The method reformulates the detection task as a question-answering framework, where URL and website content serve as the question, and the webpage's malicious classification as the answer. Using Sentence-BERT, the model extracts vector representations of questions, applies semantic clustering to refine question-answer pairs, and utilizes them as LLM prompts.
Koide \textit{et al.} \cite{koide2024chatspamdetector} developed ChatSpamDetector, an LLM-based system for phishing email detection. The system preprocesses emails by reconstructing and simplifying their content for LLM analysis. Through prompt engineering, it assigns the LLM a spam detection role and employs CoT prompting to break the analysis into sub-tasks, guiding the LLM step-by-step through the analysis process while incorporating social engineering examples.


Malware is also a common way of reconnaissance attacks. Malware will obtain sensitive information by running malicious code on the user's system\cite{yu2024maltracker}. Using LLM can effectively identify malware and prevent the leakage of sensitive user information.
Fang \textit{et al.} \cite{fang2024large} investigated LLM capabilities in defensive static analysis through a case study. Using prompt engineering, they guided GPT-4 to analyze both benign and malicious GitHub repositories, where it effectively identified malicious behaviors by recognizing characteristic patterns. Additionally, GPT-4 successfully detected malicious behaviors in the decompiled code of an Android msg-stealer virus within a comprehensive contextual framework.
Stein \textit{et al.} \cite{stein2024towards} proposed an LLM-based framework for malware detection and classification, utilizing a self-attentive mechanism to capture contextual patterns in packet sequences for distinguishing between benign and malicious traffic. Through few-shot learning, the framework effectively recognized novel malware with minimal labeled samples by generating class-specific prototypes and employing episodic training.

LLM shines in identifying hidden reconnaissance attacks due to its powerful data processing and pattern recognition capabilities. It also performs well in phishing defence due to its powerful knowledge base and logical reasoning capabilities. In addition, we also noticed that some studies have pointed out that LLM can also be used in network security education \cite{yu2024maltracker, chhetri2024exploring, is2024llmdriven}, which is also an application of LLM in defending against reconnaissance attacks.

\subsection{The Defensive Role of LLM in the Foothold Establishment Phase}


Vulnerability attacks are the most dominant methods of attack in the foothold phase, where an attacker will use specific vulnerabilities to successfully hack into a system and establish a foothold at the edge of the network to carry out subsequent attacks \cite{wang2024probabilistic}. Timely patching vulnerabilities in the system to reduce the number of exploitable vulnerabilities can significantly reduce the probability of successful intrusion by attackers. With its robust knowledge base and analysis capabilities, the LLM can efficiently complete vulnerability detection, analysis and patching and other defensive work, thus significantly increasing the speed of vulnerability remediation and reducing the workload of network defenders. All related works in this subsection are summarized in Table \ref{foothold establishment table}.

\begin{table*}
\begin{center}
\renewcommand{\arraystretch}{1.0}
\caption{Thematic Works about LLMs on Defense Against Foothold Establishment Attacks.}
\label{foothold establishment table}
\begin{tabular}{m{1.5cm} m{4.0cm} m{3.3cm} m{5.5cm} m{1.2cm}}
\toprule
\textbf{Tasks} & \textbf{Approaches} & \textbf{Targets} & \textbf{Prompt Scenario} & \textbf{Literature} \\
\midrule
\multirow{2}{1.5 cm}{\parbox[c][1.5cm][c]{1.8cm}{Vulnerability Detection}} & Code property graph, \newline Contextual learning & Detecting \newline software vulnerability & Identity, domain, in-context learning demonstrations and graph structure information & \cite{lu2024grace} \\ \cmidrule(l){2-5}
~ & Multitask self-instructed fine-tune, \newline Situational dialogue & Detecting \newline security vulnerability & Code snippets and basic task objectives & \cite{yang2024security} \\ \midrule
\multirow{3}{1.5 cm}{\parbox[c][2cm][c]{1.8cm}{Vulnerability Analysis}} & Chain templates & Building VTT mapping & Analysis objectives and vulnerability \newline description & \cite{zhang2024vtt} \\ \cmidrule(l){2-5}
~ & BERT migration learning, \newline LSTM classification &  Predicting exploitability of vulnerabilities & Analysis objectives description & \cite{yin2020apply} \\ \cmidrule(l){2-5}
~ & LLM-based parallel \newline vulnerability analysis framework & Identifying vulnerabilities & Analysis objectives description & \cite{luo2024fellmvp} \\ \midrule
\multirow{3}{1.5 cm}{\parbox[c][2.5cm][c]{1.8cm}{Vulnerability Patching}} & Prompt engineering, \newline Static Analysis & Patching smart contract Vulnerability & Role-playing, task description, external structural information, Expected Output & \cite{wang2024contracttinker} \\ \cmidrule(l){2-5}
~ & Leverage generative AI to create guiding prompts &  Patching microarchitectural side-channel vulnerabilities & Identity and task information and prompts for the type of vulnerability & \cite{tol2024zeroleak} \\ \cmidrule(l){2-5}
~ & Prompt engineering, \newline Fine-tuning & Porting hard fork patches & patching objectives description & \cite{pan2024automating} \\ \cmidrule(l){2-5}
~ & In-context learning, \newline Prompt engineering & Evaluating patches & Similar patches, bug descriptions, execution traces, failing test cases, test coverage and test patch & \cite{zhou2024leveraging} \\
\bottomrule
\end{tabular}
\end{center}
\end{table*}

\subsubsection{LLM for Vulnerability Detection}


High-speed software development technology has improved the efficiency of software development, but at the same time, it has also led to a significant increase in the number of vulnerabilities, which has brought new challenges for vulnerability detection. Recently, LLM has shown great potential in the field of vulnerability detection and has provided new ideas and methods to solve the vulnerability detection problem.


Lu \textit{et al.} \cite{lu2024grace} proposed a new approach for vulnerability detection using LLM, GRACE. The proposed method helped LLM to capture more code structure information by generating a code property graph of the detected code, and identifyed the most relevant example code to the detected code from the codebase by comparing the semantic, lexical, and syntactic similarities to provide a better demonstration in the contextual learning of LLM.
Yang \textit{et al.} \cite{yang2024security} developed MSIVD, a multitasking self-guided LLM model for vulnerability detection.They used PEFT and QLoRA techniques, fine-tuned by teacher-student dialogues, and integrated a graph neural network (GNN) to analyze the code data flow through control flow graphs. The GNN served as a lightweight adapter layer and concatenated learned embeddings at each training iteration with the hidden states of fine-tuned LLM along its last dimension. The last hidden states of an LLM encapsulated the information for all input elements before model prediction.

\subsubsection{LLM for Vulnerability Analysis}


LLM also demonstrates great potential in analysis work. Its ability to quickly identify and understand complex code structures provides defenders with timely and effective patch suggestions. This greatly shortens the cycle of vulnerability analysis and significantly improves the security and reliability of software.


Zhang \textit{et al.} \cite{zhang2024vtt} introduced VTT-LLM, a framework for mapping vulnerabilities to tactics and techniques. To enhance the LLM's reasoning capability, the authors decomposed the mapping process into four sequential steps: vulnerabilities, weaknesses, attack patterns, and ATT\&CK techniques, which are integrated as chained data during fine-tuning. 
Yin \textit{et al.} \cite{yin2020apply} proposed a framework for predicting the exploitability of vulnerabilities based on vulnerability description information. This framework applies the BERT model through transfer learning, converting tokenized wordpiece lists into embedding vectors to capture the semantic information of the wordpieces and support subsequent predictive analysis. During the fine-tuning process, the model receives the segmented vulnerability description text as input and generates token embeddings layer by layer to capture multi-level semantic information.
Luo \textit{et al.} \cite{luo2024fellmvp} introduced FELLMVP, a framework for categorizing smart contract vulnerabilities. They analyzed smart contract files to generate contract-external function-call (CEC) files that capture their semantic and structural content. These CEC files were then used to divide the dataset into eight subsets, each representing a different vulnerability type to ensure diversity. Small-batch incremental fine-tuning is performed using the LoRA method to obtain eight LLMs that specialize in identifying different vulnerability types.

In addition to the above mentioned analysis tasks, LLM has also been applied in many tasks such as penetration testing\cite{happe2023getting}, vulnerability description generation \cite{yin2024multitask}, and vulnerability localization \cite{zhang2024empirical} and so on. We believe that LLM has a broad application prospect in  vulnerability analysis, and more applications based on LLM will emerge in the future to assist security personnel to complete the work of vulnerability analysis.



\subsubsection{LLM for Vulnerability patch}


Vulnerability patching, as a key aspect of network security, has long faced the challenges of efficiency bottlenecks and resource constraints. LLM, with its powerful code understanding and generation capabilities, provides new technical ideas for automated vulnerability repair work.
Wang \textit{et al.} \cite{wang2024contracttinker} proposed a vulnerability remediation method based on the CoT mechanism, which guided LLMs to generate patches by decomposing the fixing task into a series of sub-tasks. The method also integrates static analysis techniques, including dependency analysis and program slicing, to assist LLMs in accurately locating vulnerabilities.
Tol \textit{et al.} \cite{tol2024zeroleak} presented an automated framework for patching microarchitectural side-channel vulnerabilities using LLM. The framework integrated Microwalk \cite{wichelmann2018microwalk} to locate the vulnerability and determine the cause of the vulnerability, and then utilized generative AI to craft prompts that guide LLM to deal with vulnerability. Through an iterative improvement process, LLM generated and modified patch code until the vulnerability is effectively mitigated.

In addition to generating vulnerability patches, LLMs exhibit significant potential for application in the domain of patch porting.
Pan \textit{et al.} \cite{pan2024automating} proposed a solution for automatically porting patches for hard forks using LLM, named PPatHF. They tuned the model with example from the project commit history suitable for training, and adapted the model to the porting patch task by inputting the pre- and post-patch versions of the source project in the hard fork, as well as a specific prompts.


Patch validation is also an important task in the vulnerability repair process, and this task has been facing the dual challenges of inefficiency and high cost for a long time, while the emergence of LLM provides an efficient and low-cost solution to this task.
Zhou \textit{et al.} \cite{zhou2024leveraging} evaluated the patches generated by automatic program repair using LLM without fine-tuning, utilized in-context learning, and enhanced the model's ability to judge the correctness of the patches by giving the model patch-related information.

LLM-based defence methods mainly prevent attackers from establishing the foothold through vulnerability detection, analysis and repair. Fine-tuning \cite{chen2023diversevul, guo2024outside} and prompt engineering \cite{fu2023chatgpt, nong2024chain} are currently the mainstream technical methods. Although LLM has shown good application results in these tasks, when dealing with complex and large-scale vulnerabilities, its performance is still significantly limited by the length of the input window.

\subsection{The Defensive Role of LLM in the Lateral Movement Phase}


Lateral movement is one of the most critical phases in network attack \cite{fang2022lmtracker}, enabling attackers to escalate privileges, expand system access, exfiltrate sensitive data, or compromise crucial components\cite{khoury2024jbeil}. Nowadays, the major lateral movement detection methods include IDS, anomaly detection systems, and endpoint detection and response (EDR), which mainly identify potential lateral movement activities by detecting abnormal behaviors, such as unauthorized access attempts, credentials misuse, and abnormal network traffic patterns. However, traditional detection methods rely on rules or signatures, which are difficult to cope with new types of attacks and have limitations in dealing with complex or cross-host behaviors. In contrast, LLM is able to identify anomalous behaviors in lateral movement more accurately through its pattern recognition and inference capabilities, and shows higher flexibility and adaptability, especially in the face of unknown attacks. A comprehensive summary of related works is presented in Table \ref{lateral movement table}.

\begin{table*}
\begin{center}
\caption{Thematic Works About LLMs on Defense Against Lateral Movement Attacks.}
\label{lateral movement table}
\begin{tabular}{m{1.5cm}  m{2cm}  >{\raggedright\arraybackslash}m{5cm} m{5cm} m{1.2cm}}
\toprule
\textbf{Detection Task} & \textbf{Detection Data Types} & \textbf{Data Pre-processing methods} & \textbf{Detection Data Analysis Methods} & \textbf{Literature} \\
\midrule
\multirow{2}{1.8 cm}{\parbox[c][1cm][c]{1.8cm}{Intrusion Detection}} & Vehicle network traffic data & Perform data extraction and preprocessing according to the proposed framework. & Adopt the \textbf{reasoning-followed-by-action} pipeline. & \cite{li2024ids} \\ \cmidrule(l){2-5}
~ & IoT network \newline traffic data & Use semantic extraction (BSE and USE) and input embedding. &  Analyze using the \textbf{pre-trained and fine-tuned BERT model}. & \cite{fu2024iov} \\ \midrule
\multirow{2}{1.8 cm}{\parbox[c][1.25cm][c]{1.8cm}{Anomaly Detection}} & Log data & Encode log entries into vectors and capture sequence information. & Input the \textbf{log sequence} into the \textbf{trained BERT model} to detect anomalies. & \cite{huang2023improving} \\ \cmidrule(l){2-5}
~ & Log data & Parsing logs by longest common subsequence and FT-Tree using log parsers. & Use \textbf{prompt tuning} to enable PLM to cope with different types of logging anomaly detection. & \cite{zhang2023logprompt} \\ \midrule
\multirow{1}{1.8 cm}{\parbox[c][0cm][c]{1.8cm}{EDR}} & Endpoint data & Converting endpoint data into a structured narrative form endpoint story. & Use the LLM to generate \textbf{embeddings} for each text window and detect attack behavior. & \cite{portnoy2024towards}\\
\bottomrule
\end{tabular}
\end{center}
\end{table*}

IDS are commonly used to detect lateral movement. Recently, LLMs have demonstrated the ability to perform intrusion detection across various environments, including computer networks, the Internet of Things (IoT), critical infrastructure, and cloud systems, to identify potential lateral movement behaviors\cite{kheddar2024transformers}.
Vehicle network as an emerging network is facing huge cyber threats \cite{zhang2023mitigate}. To address this issue, Fu \textit{et al.} \cite{fu2024iov} introduced IoV-BERT-IDS, a BERT-based hybrid IDS for in- and extra-vehicle networks. The system preprocesses traffic data into semantic data suitable for the BERT model through two phases. In fine-tuning, the Unidirectional Semantic Extractor (USE) converts hexadecimal strings into byte sentences by breaking packets into byte units. During pre-training, the Bidirectional Semantic Extractor (BSE) pairs neighboring packets using a sliding window, generating contextual byte sentence pairs.
Li \textit{et al.} \cite{li2024ids} developed an LLM-based IDS, IDS-Agent, which employs a structured pipeline consisting of inference, action generation, and observation updating to achieve autonomous intrusion detection. It integrates eight action spaces to simplify reasoning and improve decision accuracy. In addition, IDS-Agent utilizes structured long-term memory and external support files for inference.




Anomaly detection based on log files is one of the commonly used ways to detect lateral movement \cite{bian2021uncovering}. LLMs leverages its advanced natural language processing capabilities to parse log data and identify anomaly patterns and attack indicators\cite{le2023log}, thereby enabling effective lateral movement detection.
Huang \textit{et al.} \cite{huang2023improving} introduced HilBERT, a hierarchical transformer model tailored for system logs. The model utilizes a transformer-based log encoder to vectorize log templates and a transformer-based sequence encoder to integrate these vectors into a unified log sequence representation. Attention mechanisms are employed to capture contextual relationships and derive comprehensive insights across the sequence.
Zhang \textit{et al.} \cite{zhang2023logprompt} proposed LogPrompt, a prompt-based learning framework for log anomaly detection. It uses continuous templates with trainable vectors to adapt to diverse log structures, and a focal loss function to mitigate class imbalance between normal and anomalous logs by focusing on challenging logs.



EDR is another effective method for detecting lateral movement \cite{Smiliotopoulos2024detecting}. 
Portnoy \textit{et al.} \cite{portnoy2024towards} introduced a novel methodology for incorporating LLMs into EDR systems to improve the identification of Hands-on-Keyboard network attacks. The researchers converted raw log data into structured ``endpoint story" formats and segmented it into smaller windows. A pre-trained LLM was then employed to generate distinct embeddings for each window. These embeddings were subsequently concatenated to create an embedding sequence, which was fed into a training LLM to capture both the global context of the sequence and the inter-window relationships.

LLM-based malicious lateral movement detection is mainly achieved through the analysis of traffic data. Benefiting from LLM's powerful pattern recognition capabilities, LLM has demonstrated excellent performance in security scenarios such as IDS, anomaly detection, and EDR.

\subsection{The Research Gaps in LLM for Post-intrusion Scenarios}


Moreover, we note that the majority of current research on LLM-based defense methods primarily addresses competing and preventing network intrusion behaviors. However, a significant research gap exists in the application of LLM-based defense methods to post-intrusion scenarios, including lateral movement, data exfiltration, and post-exfiltration phases. Within the comprehensive defense lifecycle, we contend that post-intrusion network defense is of equal importance. Consequently, it is essential to expand and innovate research on LLM-based defense methods specifically designed for these scenarios, with the goal of fully leveraging the potential of LLMs in tackling post-intrusion challenges.

\section{The Application of LLM in CTI}
\label{section_6}

\begin{figure}[!t]
\centering
\includegraphics[width=0.5\textwidth]{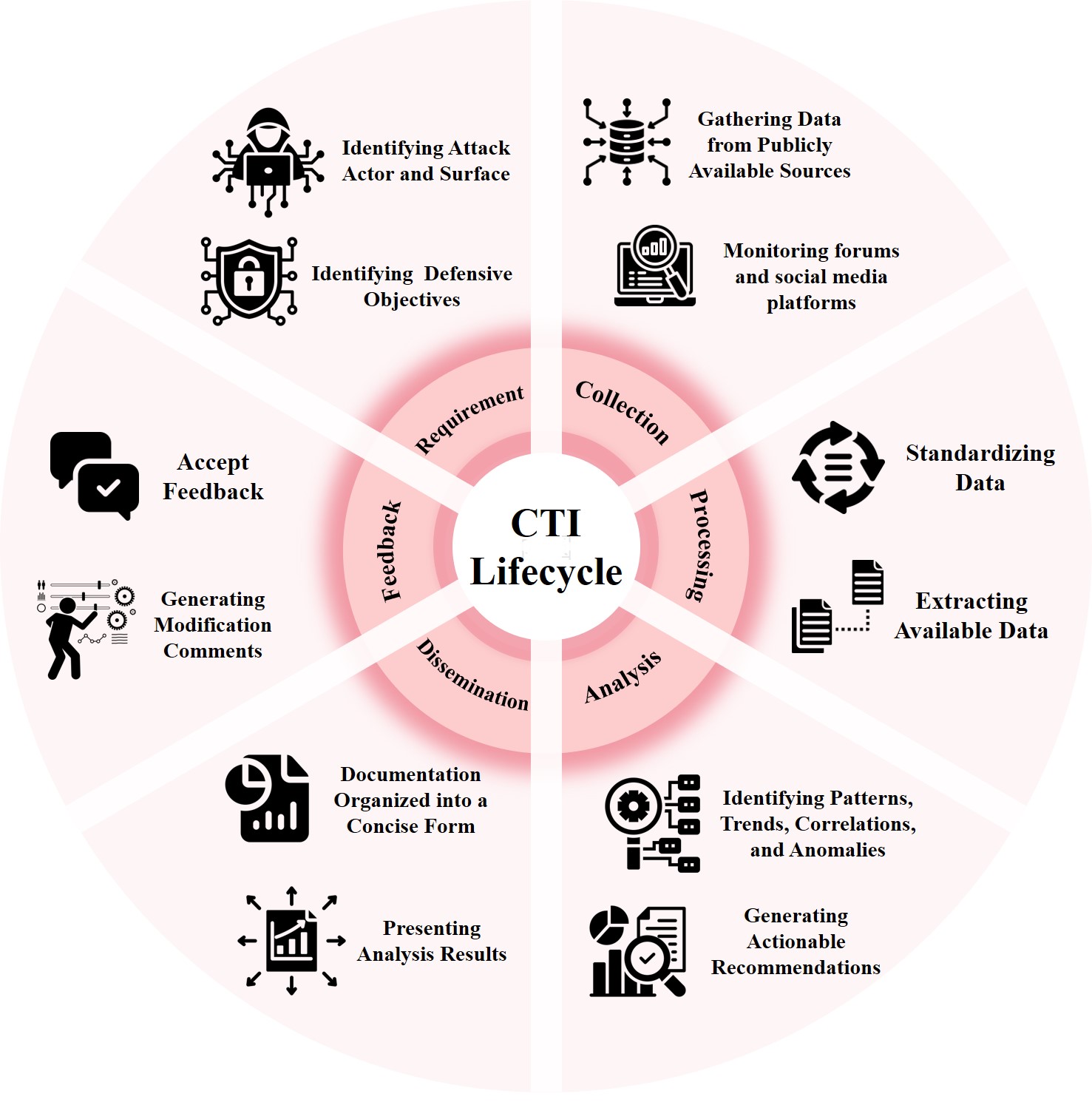}
\caption{Lifecycle of CTI.}
\label{CTI lifecycle}
\end{figure}

CTI can be defined as ``evidence-based knowledge, including context, mechanisms, indicators, implications, and actionable advice about an existing or emerging menace or hazard to assets that can be used to inform decisions regarding the subject's response to that menace or hazard"\cite{mcmillan2022definition}. It is an important part of cyber defence, the intelligence foundation of all cyber defence operations, and an important defence operation to make a defender who is in a passive situation in a cyber attack become proactive \cite{sun2023cyber}. Due to its important role in cyber defence, we also investigated the application of LLM in this defensive action.
Nova \cite{nova2022security} divided the CTI lifecycle into six phases: CTI requirements, CTI collection, CTI processing, CTI analysis, CTI dissemination, and CTI feedback, as shown as Fig. \ref{CTI lifecycle}. LLM is mainly applied in the CTI collection, processing, and analysis phases. In this section, we will introduce the application of LLM in these three phases in detail and all related works in this subsection are summarized in Table \ref{CTI table}.

\subsection{LLM for CTI Collection and Processing}


The main work in the CTI collection and processing phase is to collect data from traffic logs and publicly available sources and transform them into standardised data format for analysis\cite{nova2022security}. In this phase, there is much labour-intensive work, which significantly consumes the energy of network defenders, while LLM can efficiently complete the data collection and standardised processing tasks, which not only effectively reduces the human input but also significantly improves the efficiency of network defence work.



Collecting high-quality original CTI data that meets specific requirements often requires substantial time and effort from security personnel. The advent of LLMs offers a promising solution to mitigate this challenge.
Clairoux-Trépanier \textit{et al.} \cite{clairoux2024use} proposed system, using the GPT-3.5-turbo model to extract CTIs from cybercrime forums. This system updates data daily from cybercrime forums and provides it to the LLM, then uses ten carefully designed prompts to guide the LLM in extracting ten key variables from each forum conversation to describe CTI. 

After acquiring the raw data, it must be standardized to facilitate subsequent analysis. However, the original datasets are often unstructured, and traditional manual or rule-based processing methods are not only inefficient but also prone to errors. LLMs provide a novel approach to handle this task.
Mitra \textit{et al.} \cite{mitra2024localintel} used LLM to create a modular retrieval-augmented question-answering system, LOCALINTEL. This system generates contextualized local CTI based on the Global CTI Repository and the Local Organizational Database. The system is modular in design, and both the Global CTI Repository and the Global CTI Repository can be replaced to generate CTI that is more relevant to an organisation's needs.
Fieblinger \textit{et al.} \cite{fieblinger2024actionable} proposed a framework to extract CTI from unstructured data sources using LLM automatically. They employed the guidance framework and QLoRA fine-tuning techniques to guide LLMs in extracting CTI triples from unstructured data and subsequently organizing these triples into the structured and queryable Knowledge Graphs. The team found that the guidance framework performed better compared to prompt engineering. Also, incorporating ontology structure and a small number of examples into the prompts improved the quality of the generated CTI triples.
CTI View is a BERT-based threat entity identifier that extracts threat entities from cybersecurity report texts\cite{zhou2022cti}. In the CSKG4APT platform proposed by Ren \textit{et al.} \cite{ren2022cskg4apt}, CTI View is optimized to support the processing of cybersecurity reports in English and Chinese. The tool can automate the identification and extraction of threat entity information such as attackers, software/tools, industries, regions, and campaigns from text, providing data support for subsequent CTI analysis.

\subsection{LLM for CTI Analysing}


The main task in the CTI analysis phase is identifying patterns, trends, and anomalies from the CTI reports and getting actionable defense recommendations\cite{nova2022security}. LLM is equipped with robust information retrieval and data extraction capabilities, which can efficiently extract key information from many reports to assist in the analysis work. At the same time, LLM can use its large knowledge base to provide higher-quality defense recommendations to help network defenders make better defense decisions.


Jin \textit{et al.} \cite{jin2024crimson} presented Crimson, a system that helps LLMs convert CVE descriptions and CTI information into structured and actionable cyber defense recommendations. Crimson used a Domain-specific Embedding Model to distill complex cyber threat data into a visually and strategically insightful format and Retrieval-Aware Training (RAT) and RAT-R to enable LLMs to use contextually relevant and up-to-date cybersecurity data in their reasoning process.
Kucsván \textit{et al.} \cite{kucsvan2024inferring} developed an automated system for analyzing CTI reports and inferring threat recovery steps using LLMs. By extracting threat behavior triplets from CTI reports, the researchers employed prompt engineering to guide LLMs in deducing appropriate recovery steps.
Rahman \textit{et al.} \cite{rahman2024mining} proposed ChronoCTI, an automated pipeline for mining temporal attack patterns from CTI reports of past cyberattacks. The research team trained Roberta, a fine-tuned version of an existing LLM, using self-constructed sentences-attack techniques mapping dataset so that it can automatically identify and extract attack techniques from CTI reports.

\begin{table*}
\begin{center}
\renewcommand{\arraystretch}{1.0}
\caption{Thematic Works about LLMs on CTI.}
\label{CTI table}
\begin{tabular}{m{1.5cm}  m{4.3cm} m{3.5cm} >{\raggedright\arraybackslash}m{5.5cm}  m{1.2cm}}
\toprule
\textbf{CTI Phase} & \textbf{Task Objective} & \textbf{Processing Data Format} & \textbf{Approaches} & \textbf{Literature} \\
\midrule
\multirow{1}{3 cm}{\parbox[c][3cm][c]{2.5cm}{CTI Collection \newline and Processing}} & Extracting available CTIs from cybercrime forums & Forum information & Using \textbf{prompts} containing key CTI variables. & \cite{clairoux2024use} \\ \cmidrule(l){2-5}
~ & Generating the organization-specific threat intelligence. & Global threat databases and local knowledge databases data & Using the \textbf{modular retrieval-augmented question-answering} system. & \cite{mitra2024localintel} \\ \cmidrule(l){2-5}
~ & Extracting CTI triples from unstructured data with enhanced prompt quality. & CTI reports & Using \textbf{guidance framework} and \textbf{QLoRA fine-tuning} technique. & \cite{fieblinger2024actionable} \\ \cmidrule(l){2-5}
~ & Extracting threat entities from cybersecurity reports. & Unstructured text report collected from various sources & Text extraction and analysis of unstructured cybersecurity reports using the \textbf{BERT} model. & \cite{ren2022cskg4apt} \\ \midrule
\multirow{2}{2.5 cm}{\parbox[c][2cm][c]{2.5cm}{CTI Analysis}} & Converting CTI into structured and actionable insights. & CVEs and CTIs from diverse sources & Enhancing strategy inference in LLM using \textbf{RAT-r fine-tuning} approach. & \cite{jin2024crimson} \\ \cmidrule(l){2-5}
~ & Inferring Threat Recovery Steps from CTI Reports. & CTI reports & Guiding the LLM using the \textbf{prompt engineering} technique. & \cite{kucsvan2024inferring} \\ \cmidrule(l){2-5}
~ & Mining temporal attack patterns from CTI reports. & CTI reports & Training the \textbf{Roberta model} using self-generated training sets for the task of mining temporal patterns. & \cite{rahman2024mining}\\
\bottomrule
\end{tabular}
\end{center}
\end{table*}


Collecting, processing and analysing CTI was originally a very labour-intensive task. The introduction of LLM has revolutionised this area. The addition of LLM greatly increases the degree of automation in collecting and processing raw CTI data and also providing security personnel with effective assistance when analysing CTI reports. The use of LLM-based CTI technology effectively improves the efficiency of CTI tasks.

\section{Applications in Specific Network Scenarios}
\label{section_7}


In this section, we focus our perspective on the applications of LLM-based network security in different network scenarios, and analyze their main deployment methods and application directions from both traditional and future network perspectives.

\subsection{Traditional Network Scenarios}


The traditional network is a static network, usually with a centralized architecture, where the network has limited flexibility and scalability but is widely used. In this network scenario, LLM-based security applications are usually deployed using a centralized deployment approach, and they are generally deployed in the cloud or locally together with other security applications \cite{shen2024large}. Most of the current research also adopts this deployment approach by default, for example, the LLM-based detection systems proposed in literature \cite{fu2024iov} and \cite{huang2023improving} are deployed on local servers. Although this deployment method is less flexible, it can easily meet the high computational resource requirements of LLM and is also easy to manage. It is also worth noting that current research has been very extensive in exploring the direction of security applications of LLM in traditional network scenarios, covering key aspects such as attack detection, threat analysis, and policy generation, and LLM has demonstrated excellent performance in a number of tasks in real-world tests.

\subsection{Next-Generation Network Scenario}


Along with the technological development and demand enhancement, the concept of next-generation network has been put forward. Many types of new networks, such as IoT \cite{chi2023survey}, 6G \cite{shen2023five}, (Com)$^2$Net \cite{zhang2024com2net} and Det(Com)$^2$ \cite{zhang2024detcom2}, have emerged successively. They are characterized by high dynamics and heterogeneity, as well as multi-layered network structure \cite{zhang2024intelligent}, which brings the convenience of high adaptability, high throughput and low latency, but also brings higher network security risks than traditional networks due to the complexity of its network structure and the diversity of device access \cite{sood2023intrusion, zhu2024enabling}. On the other hand, the multi-layered network structure of next-generation network and the limited device resources also bring certain challenges to the deployment and use of traditional network security applications \cite{zhang2025moving}, and LLM-based security applications are no exception. However, in recent years, several scholars have conducted in-depth research on this challenge and proposed various solutions.

\subsubsection{Deployment}


The high demand for resources characteristic of LLM makes it difficult to be deployed on resource-limited edge devices in next-generation networks, thus limiting its usage scenarios and application performance \cite{qu2025mobile}. However, techniques such as mixture of experts \cite{zhang2025toward} and federated learning \cite{liu2024integration} can optimize the model structure and computation to achieve efficient deployment on resource-limited devices.
Zhang \textit{et al.} \cite{zhang2025toward} proposed a democratized generative AI framework using compact model strategies, where techniques such as fine-tuning, model pruning, and distillation are used to help LLM deployment on resource-constrained mobile and edge devices. 
Zhang \textit{et al.} \cite{zhang2024generative} explored the deployment issues in LLM and machine learning (ML) model-driven next-generation networks, and proposed a distributed deployment strategy that deploys LLM and traditional ML models on local or edge base stations (BSs), and each local BS owns a LLM to enhance data privacy and network scalability.
Chaoub \textit{et al.} \cite{chaoub2025mobile} proposed a hybrid deployment strategy for 6G networks. They integrated lightweight LLM sub-components into network functions (NFs) in the network for real-time tasks, and deployed complex LLM sub-components outside the NFs as standalone microservices to handle complex  analysis tasks. The two interact through service-based interfaces or fast APIs.
Xu \textit{et al.} \cite{xu2024large} proposed a split learning system for 6G networks, which realizes AI services by splitting LLMs into mobile and edge agents. The mobile side runs tiny local LLM, which is responsible for real-time sensing and local interaction, while the edge side runs huge global LLM, which performs complex reasoning and global planning. Both of them collaborate in the network to handle tasks, the mobile side can handle simple tasks independently, while the complex tasks are offloaded to the edge side for processing and then return the results for execution.

\subsubsection{Application}

The new network structure and environment of next-generation network also brings new challenges for network security defense, and the complexity of the network puts forward a new demand for intelligent network management and protection, which LLM happens to have the hope to meet.
Li \textit{et al.} \cite{li2025next}  proposed an LLM-assisted network operating system framework in which LLM management layer is integrated into the network operating system for strategy generation for service function chain deployment. In this article, the proposed NSGA2-based multi-objective LLM algorithm is innovatively used to find the optimal deployment policy. It effectively improves the intelligence and security of network management.
Liu \textit{et al.} \cite{liu2024hierarchical} pointed out that zero-trust architecture can be used to ensure the security of NGN, the article organizes the zero-trust network through micro-segmentation, and uses the Large Language Model-Enhanced Graph Diffusion (LEGD) algorithm to generate the optimal micro-segmentation, in which LLM is used to generate the dynamic filters based on the information of the network environment to reduce the algorithm search space to improve the algorithm efficiency. The article also proposes a LEGD-Adaptive Maintenance algorithm to respond to trustworthiness updates and service upgrades in the network by fine-tuning the LEGD model.
Hong \textit{et al.} \cite{Hong2024LLMTwin} proposed a framework for LLM-enabled digital twins networks (DTNs), in which LLMs will be responsible for processing multimodal data in the network. The framework utilizes LLM's own characteristics to enhance data security without affecting the network efficiency, by fine-tuning the way to load sensitive information into the LLM in DTNs, eliminating the data decryption process and reducing the possibility of data leakage, and at the same time, utilizing LLM's reversal curse characteristic \cite{lu2024rethinking} to defend against external inference attacks.
Satellite-aerial-ground integrated network (SAGIN) is also an important network architecture in NGN \cite{kong2022ergodic}, due to its heterogeneous, self-organized and dynamic characteristics, which increases the difficulty of network security protection and limits the effectiveness of traditional security methods, and LLM provides a new way to ensure the security of its network \cite{zhang2024generative}. Tang \textit{et al.} \cite{tang2024utilizing} pointed out that LLM can significantly enhance the security of SAGIN network through real-time threat detection, automated security policy formulation and dynamic security configuration. Javaid \textit{et al.} \cite{javaid2024leveraging} also proposed that in similar integrated satellite, aerial, and terrestrial networks, LLMs can traffic monitoring, malicious behavior identification, and generation of security policies to secure the network, and emphasized that LLMs can guarantee the effectiveness of LLM-enabled security measures through continuous learning.


In traditional network scenarios, LLM-based network security applications have been widely used and demonstrated excellent performance, but in next-generation network scenarios, LLM-based network security applications still face challenges in terms of resources and latency in deployment and use. However, there are now studies that have begun to study techniques such as quantization to optimize the operational efficiency of LLM in new network architectures, explore network security usage scenarios suitable for LLM, and provide feasible solutions for intelligent security protection in next-generation network environments.

\section{The Security Risk of LLM}
\label{section_8}



While LLM protects the network security, it also receives various kinds of network attacks, which may cause the weakening of LLM network defense capability or even threaten the network security of the system where it is located.

\subsection{External Threats}


LLM, as a segment of network protection, may also become the target of cyber attackers, who may attack LLM itself and paralyse the LLM-driven defence mechanism to achieve the purpose of successfully hacking into the target system. Currently, various cyber attack threats against LLMs have started to emerge, but as an emerging technology, LLMs are still under-researched in terms of their own defence. This brings great risks to users who use LLM for network defence.

\subsubsection{Prompt Injection}


Among the various attacks, prompt injection are one of the common threats to LLMs\cite{liu2024formalizing}. Such attacks enable LLMs to generate undesirable or even malicious output by embedding harmful instructions in the input\cite{huang2024survey}. For instance, an attacker may conceal malicious prompt content in logs or traffic files, directing LLMs, functioning as IDSs or EDRs, to disregard prior instructions and refrain from alerting users to the attacker’s malicious activities. 
In response to prompt injection attacks, several researchers have proposed mitigation strategies aimed at improving the robustness of LLM against such threats. These strategies aim to ensure that LLM processes input securely, executes only valid commands and resists any potential for hidden malicious content.



Chen \textit{et al.} \cite{chen2024struq} proposed a structured query approach to defend injection attacks. They used a front-end system to separate the prompt and data parts of the input, encapsulated them into a special data format, and adjusted the structured instructions to allow the LLM to accept inputs encoded in this format, thus allowing the LLM to execute only the commands in the prompt part and not the malicious commands present in the data part.
Piet \textit{et al.} \cite{piet2024jatmo} propose a method that uses fine-tuning to safeguard LLMs against prompt injection attacks. The research team exploited the fact that LLMs can only execute instructions effectively after specific instruction tuning. They used either real or self-generated datasets to fine-tune a base LLM (non-instruction-tuned) to focus more on a predefined task. This way, even if the LLM is subjected to a prompt injection attack, it will not execute the malicious instructions.

\subsubsection{Data Poisoning}


On the other hand, data poisoning attacks are also a major threat to LLM. Data poisoning attacks can occur in multiple phases of the LLM life cycle, and attackers can maliciously implant or modify some of the data in the pre-training, fine-tuning, or embedding phases of the model. They implant backdoors and loopholes in the model, which leads to a degradation of the model's performance and may even become the attacker's attack anchors to hack into the target system.
In the face of the serious threat posed by data poisoning attacks, researchers aim to enhance the robustness of the model and mitigate the impact of poisoned data on the  performance of the model as much as possible through different technical means. In the following, two defence strategies against data poisoning are presented, which propose novel solutions from different perspectives.



Li \textit{et al.} \cite{li2024purifying} employed the Kullback-Leibler divergence to measure the discrepancy between the probability distributions of the entrusted LLM and a thoroughly cleansed small language model (SLM). Their objective was to minimize the deviation between the output distribution of the integrated model and that of the SLM, thereby achieving k-Near Access-Free. This approach effectively mitigated the influence of poisoned data in the LLM on the final results while maintaining its standard performance.
Mo \textit{et al.} \cite{mo2023test} proposed a method to mitigate the impact of malicious backdoor attacks on LLMs using a small number of examples. Before user inputs are submitted to the LLM, the research team selects an example from a pool of samples designed to appropriately respond to user requirements. The selected example, which closely matches the current input task, is inserted into the input to guide the model in correctly interpreting the task instruction. 

\subsection{Inherent Risks}


Currently, LLM still has certain shortcomings in terms of performance. When responding to user needs, LLM may generate biased or incorrect replies, which can negatively impact the performance of applications integrating LLMs, and even allows the current network security system to make incorrect behaviours, which exposes the network system to huge vulnerabilities. This phenomenon is called ``misinformation" and constitutes a significant vulnerability when utilizing LLMs as a part of network defense mechanisms. The core reason for generating misinformation lies in the hallucination phenomenon of LLM. Hallucinations occur when LLMs, without truly understanding the content of the training data, use statistical patterns to fill in gaps within the training data, leading to inaccurate or misleading information.


To address the issue of misinformation in LLMs, researchers have proposed corresponding detection methods. 
Min \textit{et al.} \cite{min2023factscore} propose FACTSCORE, a framework for assessing the truthfulness of LLM-generated information.The framework subdivides LLM-generated information into atomic facts-units, which are more basic than sentences, and then, utilising the retrieval results from a knowledge source such as Wikipedia, discerns the supportiveness of each atomic facts-units, and ultimately calculates the percentage of supported atomic facts-units as the assessment score.


The hallucination phenomenon may arise due to errors or knowledge gaps in the training data used for the model\cite{huang2025survey}. To address this issue, higher-quality data can be selected during the training phase to avoid inaccuracies in the dataset. Additionally, external databases can be leveraged by modifying model parameters, injecting up-to-date knowledge, or employing RAG to provide LLM with more comprehensive knowledge. Meanwhile, the inherent limitations of the architecture and training strategies used in LLM may also contribute to the phenomenon of LLM hallucinations. However, the likelihood of LLM hallucinations can be reduced by optimising the way LLMs are trained. 
Lee \textit{et al.} \cite{lee2022factuality} proposed a Factuality-Enhanced Continued Training approach. In this approach, TOPICPREFIX, representing the topic of each training sample, is added as a prefix to improve the model's understanding of factual information. This prevents information fragmentation caused by document chunking during training. Meanwhile, the zero-masking technique is used for the front part of the sentence, and the loss function is computed only for the latter part of the sentence, which is more prone to errors, to reduce the impact of the entity misassociation problem.


Misinformation may lead to more risky matters, such as improper output handling or excessive agency issues. The improper output handling issue occurs when an erroneous output from an LLM is directly input to a downstream component or system for execution without reasonable validation or processing. In such cases, the defense system may execute incorrect operations or mislead cyber defenders, thereby increasing the likelihood of a successful attack. Conversely, the issue of excessive agency warrants serious attention. Current LLMs can invoke functions or interact with other systems via extensions. Excessive agency privileges in LLMs may pose threats to the security and integrity of network systems when errors occur.


For addressing such extended issues, on one hand, the output of LLMs can undergo secondary validation to ensure that the content is harmless to downstream components and the current system. Alternatively, the output of LLM can be restricted to allow only safe operations. For example, all database operations executed by the LLM can be limited to parameterized queries or prepared statements. On the other hand, minimizing the agency privileges granted to LLM can reduce the impact of errors on the security or integrity of the current system. Adding a ``mediator" component is also a feasible solution. The LLM can only send operation requests to the mediator, which determines whether to execute these requests, effectively intercepting unsafe calls.

\section{Open Problems and Research Directions}
\label{section_9}


Although LLM has been widely used in multiple types of cybersecurity tasks and has achieved excellent results in some of them, there are still many unresolved challenges. Meanwhile, the exploration of LLM in the field of network security is still in its early stages, and there are still many directions that can be explored. In this section, we provide a brief overview of the open problems and research directions.

\subsection{Open Problems}

\subsubsection{Scarcity of High-Quality Datasets}


The current scarcity of high-quality datasets constrains the performance optimisation of LLM on some cybersecurity tasks. On the one hand, due to the changeable forms of current attacks and the emergence of new types of attacks, it leads to the difficulty of data collection \cite{xu2025unknown}. On the other hand, the present training dataset suffers from low data accuracy and high repetitiveness, which affects the training effect and further performance optimisation of LLM \cite{croft2023data, kulsum2024case}.

\subsubsection{Input Length Limitation}


The input length limitation also affects the performance of LLM in cybersecurity tasks, especially in the field of vulnerability detection and repair. In the vulnerability detection, most of the existing solutions are limited to function-level detection, and when facing complex vulnerability detection scenarios involving cross-function or cross-class vulnerabilities, LLM is unable to handle complete code fragment information, resulting in a significant degradation of its detection performance \cite{zhou2024comparison}. Similarly, in vulnerability repair, the limited context window of LLM cannot accommodate the complete semantic information of a large codebase and the related project background information, which makes it difficult to read enough repair help information, resulting in poor repair results \cite{kulsum2024case}.

\subsubsection{Targeted Attack Threats}


LLM is vulnerable to targeted attacks when performing detection and collection of external threat information. For instance, during phishing email detection, the email may contain injection attacks\cite{koide2024chatphishdetector}. Or during CTI collection, LLM may be attacked by toxic data due to the complexity of raw intelligence data sources, including unreliable sources such as cybercrime forums and the dark web.

\subsubsection{Problem of Delayed Inference}


In scenarios such as honeypot systems and intrusion detection systems, which have the requirements on system response speed, LLM currently still has the problem of slow real-time response speed, and we think that most of the researches are now mainly focusing on the optimisation of the LLM-base honeypot deception effect and the optimisation of the precision of attack detection, and less consideration is given to the real-time response capability of the LLM-based defence system \cite{vo2024apelid}.

\subsubsection{Black-Box LLMs}


Most current studies mostly use black-box LLMs \cite{pearce2023examining}, such as the OpenAI GPT series. Although these LLMs have excellent performance, the experimental test samples may have been covered by the pre-training data due to the opacity of the training data, leading to doubts about the reproducibility and reliability of the high repair success rate obtained from the experiments in real scenarios.

\subsection{Research Directions}

\subsubsection{Development of Open-Source and Transparent Datasets}


The development of open-source and transparent high-quality datasets can continue to alleviate the current problem of dataset scarcity on the one hand, and on the other hand, mitigate the risk of uncertainty in experimental results due to black-box LLM.

\subsubsection{Breaking Through the Input Length Limitation}


The input length of LLM limits its performance in vulnerability detection and remediation, we think that with the development of LLM technology, the enhancement of the LLM input window capacity may alleviate this problem, on the other hand, current techniques such as code property graph \cite{lu2024grace} or GNN \cite{yang2024security} can also be used to compress the information to alleviate the impact of the input length limitation.

\subsubsection{Designing Defence Mechanisms Against Pollution Attacks}


LLMs frequently interact with unsafe external data in cybersecurity tasks, which may contain pollution inputs such as injection attacks. Such exposure risks degrading model performance or even co-opting LLMs as attack vectors. We think that developing data filtering or isolation techniques to shield LLMs from adversarial contamination is both imperative and a pivotal future research direction.

\subsubsection{Semantic Data Transformation}


As LLM is applied to a wider range of cybersecurity tasks in the future, the data that LLM needs to process will be more complex and diverse, and may not be semantic data suitable for LLM processing. We think that future research needs to allow LLM to cope with more diverse data processing needs, which can be accomplished through data preprocessing, prompt engineering, or domain-specific data training. Future research also focuses on constructing efficient data transformation mechanisms that preserve the critical features and contextual relationships of the original data while reformatting it into information-rich semantic representations.

\subsubsection{Enhancing Interpretability}


Improving the interpretability of LLM analysis results is also an important research direction in the future. In most attack threat detection work, improving interpretability can help security personnel understand the detection results more intuitively, and improve the transparency and trust of the system. It also helps subsequent research to optimise the performance of LLM.

\subsubsection{Expanding Application Coverage}


With the concept of next-generation network, a number of new network architectures have emerged to meet various new demands. While improving network performance, these architectures also bring more attack surfaces and higher security risks, while the complex and changing network structure also increases the difficulty of network protection. LLM with intelligent and adaptive features provides a new solution idea for future network defense. Expanding LLM-based network security solutions from traditional network scenarios to next-generation network scenarios, and studying the deployment and application of network defense solutions in network scenarios such as 6G, IoT, and SAGIN, etc. will be a mainstream research direction in the future.

\section{Conclusion}
\label{section_10}




In this survey, we not only explore the defensive role of LLM in the various life cycles of cyber attacks, but also point out the obvious research gaps in the post-intrusion scenario. Through the analysis of relevant literature, we clearly point out the huge application potential of LLM in network security. Although there are already many studies using LLM to accomplish cybersecurity tasks, it should be noted that there are still many unresolved issues and challenges in LLM-based applications. Based on the current status of LLM applications in cybersecurity, we have also listed some possible future research directions. We hope that through this survey, we can provide a systematic thinking and reference framework for future research on the application of LLM in cybersecurity.

\bibliographystyle{IEEEtran}
\bibliography{ref}



\begin{IEEEbiography}[{\includegraphics[width=1in,height=1.25in,clip,keepaspectratio]{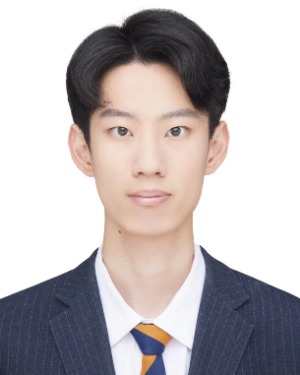}}]{Shuang Tian}
received the B.Eng. degree in computer science and technology from China University of Geosciences Beijing, Beijing, China, in 2024. He is currently working toward the M.Eng. degree with the School of Software Engineering, Beijing Jiaotong University, Beijing, China.
\end{IEEEbiography}

\begin{IEEEbiography}[{\includegraphics[width=1in,height=1.25in,clip,keepaspectratio]{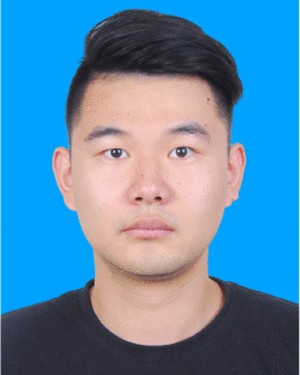}}]{Tao Zhang}
(Member, IEEE) received the B.S. degree in Internet of Things engineering from the Beijing University of Posts and Telecommunications (BUPT) and the Queen Mary University of London in 2018, and the Ph.D. degree in computer science and technology from BUPT in 2023. He is currently an Associate Professor with the School of Cyberspace Science and Technology, Beijing Jiaotong University. His publications include ESI highly cited paper and well-archived international journals and proceedings, such as IEEE COMST, JSAC, TIFS, TDSC, TMC, TITS, TCCN and TII etc. His research interests include network security, moving target defense, and federated learning. He has served as the guest editor for Electronics and Chinese Journal of Network and Information Security, and the TPC chair and a PC member for some international conferences and workshops. He was a recipient of the Best Paper Award from NaNA 2018, IWCMC 2021, DIONE 2024, and ICA3PP 2024, and a recipient of Outstanding Paper Award from iThings 2023, and SmartCity 2024. His Ph.D. thesis was awarded the Outstanding Doctoral Dissertation by BUPT in 2023.
\end{IEEEbiography}

\begin{IEEEbiography}[{\includegraphics[width=1in,height=1.25in,clip,keepaspectratio]{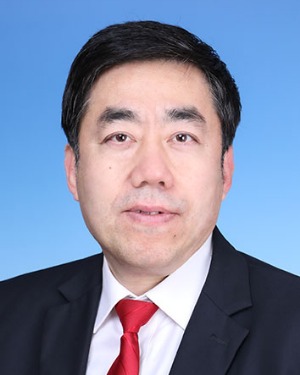}}]{Jiqiang Liu}
(Senior Member, IEEE) received the Ph.D. degree from Beijing Normal University in 1999. He is currently a Full Professor and the Dean of the School of Cyberspace Science and Technology, Beijing Jiaotong University. He has authored or coauthored over 200 publications. In recent years, he has been mainly engaged in research on trusted computing, privacy protection, and cloud computing security.
\end{IEEEbiography}

\begin{IEEEbiography}[{\includegraphics[width=1in,height=1.25in,clip,keepaspectratio]{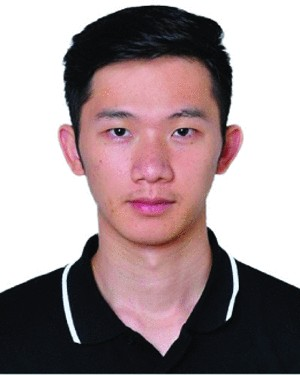}}]{Jiacheng Wang}
is the postdoctoral research fellow in the College of Computing and Data Science, Nanyang Technological University, Singapore. Prior to that, he received the Ph.D. degree in School of Communications and Information Engineering, Chongqing University of Posts and Telecommunications, Chongqing, China. His research interests include wireless sensing, generative artificial intelligence, and semantic communications
\end{IEEEbiography}

\begin{IEEEbiography}[{\includegraphics[width=1in,height=1.25in,clip,keepaspectratio]{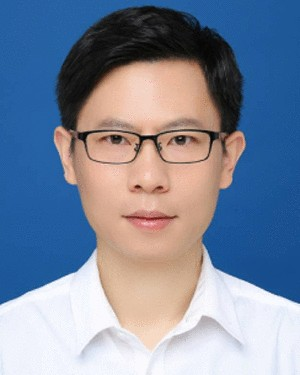}}]{Xuangou Wu}
received the Ph.D. degree from the School of Computer Science and Technology, University of Science and Technology of China, Hefei, China, in 2013. He is currently a Full Professor and the Dean of the School of Computer Science and Technology, University of Anhui Technology, Maanshan, China. His research interests include Intelligent Internet of Things, Network Security, and Privacy Protection.
\end{IEEEbiography}

\begin{IEEEbiography}[{\includegraphics[width=1in,height=1.25in,clip,keepaspectratio]{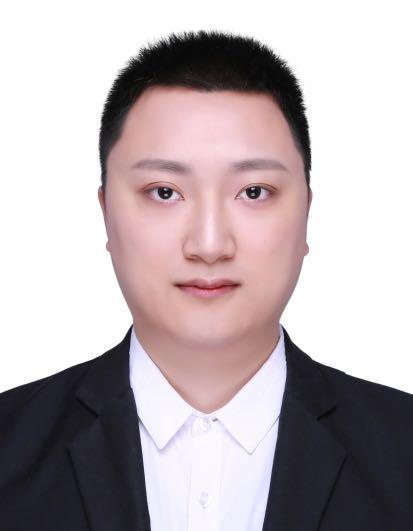}}]{Xiaoqiang Zhu}
(M'23) received the Ph.D. degree in software engineering from Tianjin University, China, in 2022, and the M.S. degree in computer science from Dalian University of Technology, China, in 2018. He served as a joint Ph.D. student at ETH Zurich, Switzerland, supported by the China Scholarship Council in 2021. He is currently an Assistant Professor (Lecturer) with the School of Cyberspace Science and Technology, Beijing Jiaotong University, China. He has published scientific papers in international journals, such as IEEE COMST, TMC, TNSE, etc..
\end{IEEEbiography}

\begin{IEEEbiography}[{\includegraphics[width=1in,height=1.25in,clip,keepaspectratio]{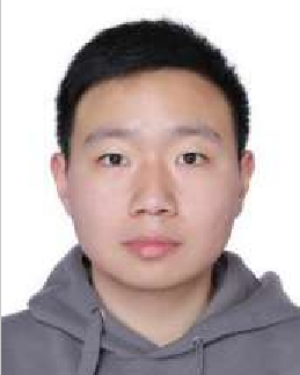}}]{Ruichen Zhang}
(Member, IEEE) is currently working as a PostDoctoral Research Fellow with the College of Computing and Data Science, Nanyang Technological University (NTU), Singapore. He received the B.E. degree from Henan University (HENU), China, in 2018, and the Ph.D. degree from Beijing Jiaotong University (BJTU), China, in 2023. In 2024, he was a Visiting Scholar with the College of Information and Communication Engineering, Sungkyunkwan University, Suwon, South Korea. His research interests include LLM-empowered networking, reinforcement learning-enabled wireless communication, generative AI models, and heterogeneous networks.
\end{IEEEbiography}

\begin{IEEEbiography}[{\includegraphics[width=1in,height=1.25in,clip,keepaspectratio]{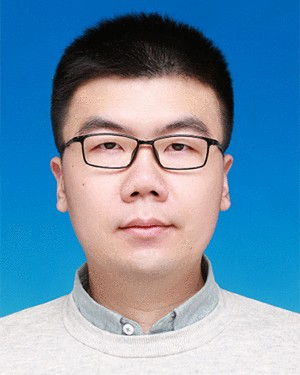}}]{Weiting Zhang}
(Member, IEEE) received the PhD degree in communication and information systems with Beijing Jiaotong University, Beijing, China, in 2021. From 2019 to 2020, he was a visiting PhD student with the Department of Electrical and Computer Engineering, University of Waterloo, Canada. Starting from December 2021, he works as an associate professor with the School of Electronic and Information Engineering, Beijing Jiaotong University. His research interests include industrial Internet of Things, federated learning and edge intelligence.
\end{IEEEbiography}

\begin{IEEEbiography}[{\includegraphics[width=1in,height=1.25in,clip,keepaspectratio]{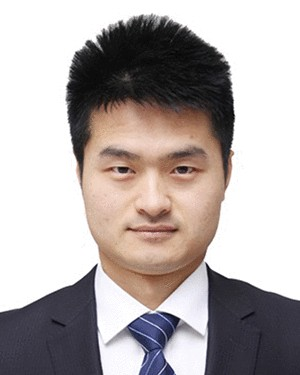}}]{Zhenhui Yuan}
(Senior Member, IEEE) received the B.Eng. degree in software engineering with Wuhan University, Wuhan, China, in 2008, and the Ph.D. degree in electronic engineering from Dublin City University, Dublin, Ireland, in 2012. He is currently an Assistant Professor with the University of Warwick. He was the Lead Guest Editor in IEEE NETWORK and IEEE INTERNET OF THINGS JOURNAL. He is also the founding Chair of VeSUS (6G-empowered Robotic Vehicles for Sustainable Development) Workshop.
\end{IEEEbiography}

\begin{IEEEbiography}[{\includegraphics[width=1in,height=1.25in,clip,keepaspectratio]{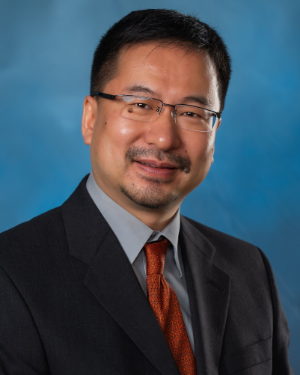}}]{Shiwen Mao}
(Fellow, IEEE) is a Professor and Earle C. Williams Eminent Scholar and Director of the Wireless Engineering Research and Education Center at Auburn University. Dr. Mao’s research interest includes wireless networks, multimedia communications, and smart grid. He is the editor-in-chief of IEEE Transactions on Cognitive Communications and Networking and a member-atlarge on the Board of Governors of IEEE Communications Society. He received the IEEE ComSoc MMTC Outstanding Researcher Award in 2023, and the SEC 2023 Faculty Achievement Award for Auburn.
\end{IEEEbiography}

\begin{IEEEbiography}[{\includegraphics[width=1in,height=1.25in,clip,keepaspectratio]{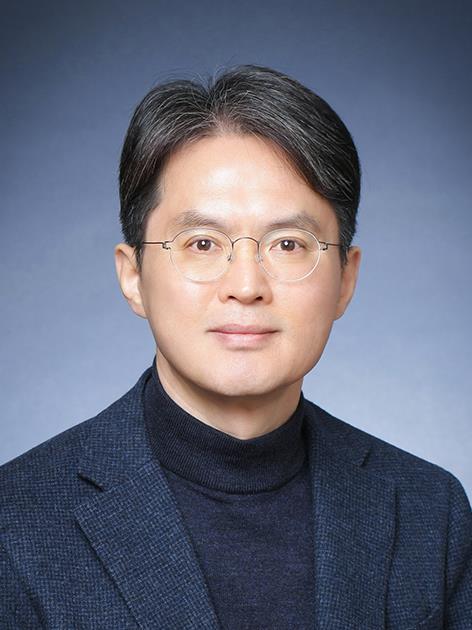}}]{Dong In Kim}
(Life Fellow, IEEE) received the Ph.D. degree in electrical engineering from the University of Southern California, Los Angeles, CA, USA, in 1990.  He is currently a Distinguished Professor with the College of Information and Communication Engineering, Sungkyunkwan University, Suwon, South Korea. He is a Fellow of the Korean Academy of Science and Technology and a Life Member of the National Academy of Engineering of Korea. He received several research awards, including the 2023 IEEE ComSoc Best Survey Paper Award and the 2022 IEEE Best Land Transportation Paper Award.
\end{IEEEbiography}

\vspace{11pt}

\vfill

\end{CJK}
\end{document}